\documentclass[twocolumn,prd,aps,showpacs,superscriptaddress]{revtex4}
\usepackage{amssymb}
\usepackage{epsf}
\usepackage{amsmath}
\usepackage{amsfonts}

\begin{document}

\title{Slowly rotating fluid balls
of Petrov type D}
\author{Michael Bradley\footnote{electronic address: michael.bradley@physics.umu.se}} 
\affiliation{Department of Physics, Ume\aa\ University, SE-901 87 Ume\aa, Sweden}

\author{Daniel Eriksson\footnote{electronic address: daniel.eriksson@physics.umu.se}} 
\affiliation{Department of Physics, Ume\aa\ University, SE-901 87 Ume\aa, Sweden}

\author{Gyula Fodor\footnote{electronic address: gfodor@rmki.kfki.hu}} 
\affiliation{KFKI Research Institute for Particle and Nuclear Physics,
H-1525, Budapest 114, P.O.B. 49, Hungary}

\author{Istv\'an R\'acz\footnote{electronic address: iracz@sunserv.kfki.hu}}
\affiliation{KFKI Research Institute for Particle and Nuclear Physics,
H-1525, Budapest 114, P.O.B. 49, Hungary}

\begin{abstract}

The second order perturbative field equations for slowly and rigidly
rotating perfect fluid balls of Petrov type D are solved numerically.
It is found that all the slowly and rigidly rotating perfect fluid
balls up to second order, irrespective of Petrov type, may be matched
to a possibly non-asymptotically flat stationary axisymmetric vacuum
exterior. The Petrov type D interior solutions are characterized by
five integration constants, corresponding to density and pressure of
the zeroth order configuration, the magnitude of the vorticity, one
more second order constant and an independent spherically symmetric
second order small perturbation of the central pressure.  A
four-dimensional subspace of this five-dimensional parameter space is
identified for which the solutions can be matched to an asymptotically
flat exterior vacuum region.  Hence these solutions are completely
determined by the spherical configuration and the magnitude of the
vorticity.  The physical properties like equations of state, shapes
and speeds of sound are determined for a number of solutions.

\end{abstract}

\pacs{04.40.Dg, 04.25.-g, 04.20.-q}

\maketitle

\section{Introduction}

In \cite{Hartle} a second order formalism for slowly and rigidly
rotating stars was developed by Hartle.  This formalism was in
\cite{HartleThorne} applied to rotating white dwarfs and neutron stars
using the Harrison-Wheeler and Tsuruta Cameron $V_{\gamma}$ equations
of state and in \cite{ChandrasekharMiller} to the case with constant
energy density. In \cite{Bertietal} the second order formalism is
compared with numerical solutions of the full Einstein equations.  For
a review of relativistic rotating stars see \cite{Stergioulas}. In
\cite{Cabezas} global models for slowly rotating bodies in the
post-Minkowskian approximation are treated.  In a recent paper
\cite{VeraMarsMacCallum} second order perturbation theory for the
matching of general stationary axisymmetric bodies to an
asymptotically flat vacuum has been put on a more solid mathematical
ground and the exterior metric is determined to second
order.

In this paper we use the Hartle formalism to study perfect fluids of
Petrov type D. This condition will be used instead of an equation of
state. It was shown in \cite{FodPer} that physically realistic
rotating fluid balls cannot be of Petrov types II, III, N or 0 so the
only possible cases are of Petrov types D or I. Hence it is of
interest to closer study the properties of Petrov type D solutions,
being the only possible algebraically special solutions. Also, since
in the non-rotating spherically symmetric case all interior solutions
are of Petrov type D or 0, one might hope to find physically
interesting interior solutions of Petrov type D also in the
axisymmetric case, at least for slow rotation.  However, the
quadrupole moment of the rotating configuration will typically deviate
from that given by the Kerr metric and hence its exterior metric
cannot be Kerr \cite{Bertietal}. It is easily verified that such an
exterior metric is not of Petrov type D.

The field equations to second order in the small rotational parameter
$\Omega$ will be solved numerically using fourth order Runge-Kutta.
The system reduces to a closed subsystem of six first order
differential equations. There are also two more differential
equations for two further dependent variables which do not appear
in this closed subsystem.  Assuming
regularity at the centre the solutions of this closed subsystem depend
on four constants of integration, corresponding to zeroth order
central density and pressure, the magnitude of the angular velocity
and one more second order small constant.  Due to scaling invariances
we need only consider a two dimensional subspace of the solution
space.  The solutions are then matched to a second order axisymmetric
vacuum solution using the Darmois-Israel procedure \cite{Darmois,Israel}.  
This metric includes the general second order asymptotically
flat stationary axisymmetric vacuum solution as a special case. The
interior solutions that can be matched to this vacuum form a three
dimensional subspace of the space of solutions.  One more freely
specifiable parameter, associated with an independent spherical
symmetric second order small change of the central pressure, is
obtained from the solution of the two remaining equations.  Hence the
rotating configuration for the asymptotically flat subclass is
determined by the spherically symmetric configuration (including a
possible second order change of the central pressure) and the
magnitude of the angular velocity.

The paper is organized as follows: In section \ref{Preliminaries} the
method is briefly described and the field equations are presented,
along with the Petrov type D condition. Finally, the second order
vacuum metric is given.  The matching procedure is described in
section \ref{Matching} and the integration constants for the vacuum
solution are solved for in terms of the values of the interior
solution on the matching surface. In section \ref{Newvariables} the
equations are rewritten in a form suitable for numerical
integration. The results of the numerical runs are given in section
\ref{Numericalsolutions}. First the program is checked against the
exact Wahlquist solution and then the subset for which the solutions
are asymptotically flat is determined.  Properties like shape,
equation of state and speed of sound are then determined for a number
of solutions.

\section{Preliminaries}\label{Preliminaries}

To second order the metric of a slowly rotating axisymmetric object,
both in the interior fluid region and the outside vacuum region, can
be written as
\begin{eqnarray}\label{hartlemetric}
ds^{2}  &=&(1+2h)A^2dt^{2}-(1+2m)\frac{1}{B^2}dr^{2}-\nonumber\\
& &  (1+2k)r^{2}\left[  d\theta^{2}+\sin^{2}\theta\left(
d\varphi-\omega dt\right)  ^{2}\right]  \ , \label{ds}
\end{eqnarray}
where $\omega$ is first order and $h$, $m$ are $k$ are second order in
the rotational parameter \cite{Hartle}.  The requirements of
regularity at the centre and asymptotic flatness imply that the first
order function $\omega$ depends on $r$ only.  The second order
functions $h,m,k$ can be given as
\begin{eqnarray}\nonumber
h&=&h_0+h_2P_2(\cos\theta)\\\nonumber
m&=&m_0+m_2P_2(\cos\theta)\\
k&=&k_2P_2(\cos\theta)
\end{eqnarray}
where $h_0,m_0$ and $h_2,m_2,k_2$ are functions of $r$ only and
$P_2(x)=\frac{1}{2}(3x^2-1)$ is the second order Legendre polynomial.
This result follows from reflection symmetry in the equatorial plane,
from that the equations for $h$, $m$ and $k$ separate with the
ans\"atze $h=\sum_{i=0}^{\infty}h_i(r)P_i(\cos\theta)$ etc., and from
the fact that there are no inhomogeneous terms containing $\omega$ in
the equations for $h_i$, $k_i$ and $m_i$ for $i>2$. For more details
see \cite{Hartle}.

The matching of the two spacetime regions happens via the application
of a coordinate transformation $\varphi \rightarrow \varphi+\Omega t$
in the fluid region. In addition to this, we can also rescale the
interior time coordinate first by a constant $c_4$ while matching the
spherical zeroth order solutions, and then later by a second order
small constant when doing the matching of the corresponding rotating
spacetimes. These yield then the coordinate transformation $t
\rightarrow c_4(1+c_3)t$.  The first of these coordinate
transformations says that the inner fluid region rotates with respect
to the distant stationary observers with angular velocity $\Omega$.
This parameter $\Omega$ is considered to be the small expansion
parameter with respect to which $\omega$ is first order and the other
corrections $h,m,k$ are second order.

The matter content of the interior is modelled by a perfect fluid
\begin{displaymath}
T_{ab}=(\rho+p)u_{a}u_{b}-pg_{ab} \ .
\end{displaymath}
The coordinate system used in (\ref{hartlemetric}) is assumed to be
comoving with the fluid, i.e. the 4-velocity is assumed to possess the
form
\begin{displaymath}
u^a=(1/\sqrt{g_{00}},0,0,0)=((1-h)/A,0,0,0)
\end{displaymath}
which also implies that the shear of the fluid is zero so it rotates
rigidly.

\subsection{The field equations}\label{sectionfieldeq}

The basic field equations relevant to various orders can be listed as
follows \cite{Fodor}.  If no equation of state is specified then the
only equation one gets to zeroth order of the rotational parameter is
the pressure isotropy condition $G^1\,_1=G^2\,_2$ which reads as
\begin{equation}\label{G110}
B\frac{d^{2}A}{dr^{2}}
+\frac{d (rA)}{dr}\frac{d (B/r)}{dr}
+\frac{A}{r^2 B}=0
\ .
\end{equation}
Making use of $G^0\,_0=T^0\,_0$ and $G^1\,_1=T^1\,_1$ the energy
density and pressure of the non-rotating configuration reads as
\begin{eqnarray}
\rho_0 &=&\frac{1}{r^{2}}\left[  1-\frac{d(rB^2)}{dr}\right]  \ ,\label{mu}\\
p_0 &=&\frac{1}{r^{2}}\left[  \frac{B^2}{A^2}\frac{d(rA^2)}{dr}-1\right]
\ .\label{pp}
\end{eqnarray}
To first order in the rotation parameter the only relation follows
from $G^3\,_0=0$
\begin{equation}\label{G30}
\frac{d}{dr}\left(  r^{4}\frac{B}{A}\frac{d\omega}{dr}\right)
+4r^{3}\omega\frac{d}{dr}\left(  \frac{B}{A}\right)  
=0\ . 
\end{equation}
The second order Einstein equations yield the following four conditions.
 From $G^1\,_2=0$ one gets 
\begin{equation}\label{G12}
r\frac{d}{dr}\left(  h_{2}+k_{2}\right)  
+r\left(  h_{2}-m_{2}\right)\frac{1}{A}
\frac{dA}{dr}-  h_{2}-m_{2}  =0\ . 
\end{equation}
The pressure isotropy condition in the angular directions,
$G^2\,_2=G^3\,_3$, gives
\begin{equation}\label{G22}
6\left(  h_{2}+m_{2}\right) -r^4 \frac{B^2}{A^2} 
\left( \frac{d\omega}{dr}\right)^2
+4r^3\omega^2\frac{B}{A}\frac{d}{dr}\left( \frac{B}{A}\right)=0 \ .
\end{equation}
The equality of the pressure in the angular and radial directions,
i.e.\ $G^1\,_1=G^2\,_2$ gives two equations.  After eliminating the
derivative of $h_2$ using (\ref{G12}) one obtains from the
$P_2(\cos\theta)$ part
\begin{eqnarray}\nonumber
&&2r\frac{B^2}{A}\frac{d A}{dr}\left(r\frac{d k_2}{dr}-m_2\right)
-2r^2Bh_2\frac{d }{dr}\left(\frac{B}{r}\right)
+ \\\label{G1122}
&&\ \ \ \ \ \ \ 
m_2-4k_2-5h_2-\frac{1}{3}r^4\frac{B^2}{A^2}\left( 
\frac{d\omega}{dr}\right)^2=0 \ ,
\end{eqnarray}
while the $\theta$-independent part takes the form
\begin{eqnarray}\nonumber
&& 
6r^3B\frac{d}{dr}\left(\frac{1}{r}A^2B\frac{d h_0}{dr}\right)
-3B^2\frac{d(r^2A^2)}{dr}\frac{d m_0}{dr}+ \\\nonumber
&&12A^2m_0-3r^4B^2\left( \frac{d\omega}{dr}\right)^2
+4r^3\omega^2AB\frac{d}{dr}\left( \frac{B}{A}\right)=0 \ .\\\label{G1120}
\end{eqnarray}

The energy density function can be decomposed as $\rho=\rho_0+\rho_2$,
where $\rho_2=\rho_{20}+\rho_{22}P_2(\cos\theta)$ and $\rho_{20}$ and
$\rho_{22}$ are second order small functions of the coordinate $r$
given as
\begin{eqnarray}\nonumber
\rho_{20}=\frac{B}{6r^2A}\left[
8r^3\omega^2\frac{d}{dr}\left(\frac{B}{A}\right)
+12\frac{A}{B}\frac{d}{dr}\left(r B^2m_0\right) \right.\\\label{r20}
\left.
-r^4\frac{B}{A}\left(\frac{d\omega}{dr}\right)^2
\right]
\end{eqnarray}
and
\begin{equation}
\rho_{22}=-\frac{2(3A^2h_2+r^2\omega^2)}{3r^3A\frac{dA}{dr}}
\left[1-B^2+r^2\frac{d}{dr}\left(B\frac{dB}{dr}\right)\right] \ .
\end{equation}
The analogous decomposition of the pressure is defined by
$p=p_0+p_2=p_0+p_{20}+p_{22}P_2(\cos\theta)$, where
\begin{equation}\label{p20}
p_{20} = \frac {B^2}{6r^2A^2} \left[ 12rA^2\frac{d h_0}{dr}-
12m_0\frac{d}{dr}(rA^{2})+r^4\left(\frac{d\omega}{dr}\right)^2\right] 
\end{equation}
and
\begin{equation}\label{p22}
p_{22} =\frac{2 B}{3 r A}\left(3 A^2 h_2 + r^2 \omega^2\right) \frac{d}{d r} 
\left( \frac{B}{A}\right) \, .
\end{equation}

The existence of a barotropic equation of state $\rho=\rho(p)$ is
equivalent to
\begin{equation}
\frac{\partial \rho}{\partial \theta}\frac{\partial p}{\partial r}-
\frac{\partial p}{\partial \theta}\frac{\partial \rho}{\partial r}=0\ ,
\end{equation}
which is a geometric condition ensuring the coincidence of the
constant pressure and density surfaces.  Substituting the
decompositions $\rho=\rho_0+\rho_{20}+\rho_{22}P_2(\cos\theta)$ and
$p=p_0+p_{20}+p_{22}P_2(\cos\theta)$ into this relation yields, up to
second order,
\begin{equation}
\rho_{22}\frac{d p_0}{d r}=
p_{22}\frac{d\rho_0}{d r} \ , \label{exeos}
\end{equation}
which is identically satisfied in virtue of the above field equations.

It seems to be plausible to require the equation of state to be
independent of the angular velocity. This condition reads as
\begin{equation}
\frac{\rho_2}{p_2}=
\frac{d\rho_0}{d p_0} \ .
\end{equation}
The $\theta$-dependent part of this relation is equivalent to
(\ref{exeos}), while the spherically symmetric part gives the relation
\begin{equation}
\rho_{20}\frac{\partial p_0}{\partial r}=
p_{20}\frac{\partial \rho_0}{\partial r} \ .
\end{equation}
Then by the substitution of the expressions $\rho_{20}$ and $p_{20}$,
given by (\ref{r20}) and (\ref{p20}), together with $\rho_0$ and $p_0$
for the zeroth order pressure and density, given by (\ref{pp}), one
gets
\begin{eqnarray}\nonumber
&&24rA^4\left[r^2\frac{d^2B^2}{dr^2}+2(1-B^2)\right]\frac{d h_0}{d r}+
24rAm_0\frac{dA}{dr}\times
\\\nonumber
&&\left[4A^2(B^2-1)-4r^2AB\frac{dA}{dr}\frac{dB}{dr}
-4r^2A^2B\frac{d^2 B}{dr^2}\right]+\\\nonumber
&& 8r^3A^5\frac{d A}{dr}\frac{d}{dr}\left(\frac{B^2}{A^2}\right)
\left[3\frac{d m_0}{dr}+\frac{r^2\omega^2}{B^2}\frac{d}{dr}\left(\frac{B^2}{A^2}\right)\right]+
\\\nonumber
&&A^2\left[2r^2\frac{d^2B^2}{dr^2}-2r^2A\frac{dA}{dr}\frac{d}{dr}
\left(\frac{B^2}{A^2}\right)+4(1-B^2)\right]\times\\\label{uncheos}
&&\left[r^4\left(\frac{d\omega}{dr}\right)^2-12A^2m_0\right]=0 \ .
\end{eqnarray}

\subsection{The Petrov type of slowly rotating fluids}

The spherically symmetric field equation (\ref{G110}) is usually
complemented by a choice of an equation of state for the fluid.  Since
spherically symmetric static spacetimes are always algebraically
special, we do not assume any special equation of state for the
non-rotating base solution.  As we will see shortly, the deviation
from algebraically special cases can arise first when considering the
second order terms in the rotational parameter.  Here we require the
interior solution to remain Petrov type D for slow rotation, thereby
completing the system of field equations (\ref{G110}), (\ref{G30}),
(\ref{G12}), (\ref{G22}), (\ref{G1122}), (\ref{G1120}) and
(\ref{uncheos}) by a further condition, which in some sense plays the
role of an equation of state.

In order to calculate the Petrov type we need a suitable null
tetrad. Up to second order an orthonormal tetrad can be given as
\begin{eqnarray}
e^\mu_0&=&\left(\frac{1}{A}\left(1
+\frac{\omega^2r^2}{2A^2}\sin^2\theta-h\right)
,0,0,0\right) \nonumber\\
e^\mu_1&=&\left(0,B(1-m),0,0\right) \nonumber\\
e^\mu_2&=&\left(0,0,\frac{1}{r}(1-k),0\right) \nonumber\\
e^\mu_3&=&\left(\frac{r\omega}{A^2}\sin\theta,0,0,
\frac{1}{r\sin\theta}\left(-1
+\frac{\omega^2r^2}{2A^2}\sin^2\theta+k\right)
\right) \ . \nonumber
\end{eqnarray}
 From this we form the null tetrad by the relations 
\begin{eqnarray}
\sqrt{2}\,l^\mu&=&e^\mu_0+e^\mu_3 \nonumber\\
\sqrt{2}\,k^\mu&=&e^\mu_0-e^\mu_3 \nonumber\\
\sqrt{2}\,m^\mu&=&e^\mu_1+i e^\mu_2 \ .\nonumber
\end{eqnarray}
Then the components of the Weyl spinor are given as
\begin{eqnarray}
&&\Psi_0=C_{abcd}k^am^bk^cm^d \ \ \ \ 
\Psi_3=C_{abcd}k^al^b\bar m^cl^d
\nonumber\\
&&\Psi_1=C_{abcd}k^al^bk^cm^d \ \ \ \ \ \ 
\Psi_4=C_{abcd}\bar m^al^b\bar m^cl^d
\nonumber\\
&&\Psi_2=C_{abcd}k^am^b\bar m^cl^d \ . 
\nonumber
\end{eqnarray}
Since $\Psi_1=0$ and $\Psi_3=0$ (even in case of fast rotation) the
Petrov type is determined by the multiplicities of the roots of the
algebraic equation for the complex number $a$
\begin{equation}
\Psi_0+6\Psi_2a^2+\Psi_4a^4=0 \ .
\end{equation}
The Petrov type is D if there are two double roots, i.e.
\begin{equation}
9\Psi_2^2=\Psi_0\Psi_4 \ . \label{petcond1}
\end{equation}
We note that the Petrov type can also be D if $\Psi_0=\Psi_4=0$ and
$\Psi_2\not=0$ but since then the equation of state can be shown to be
$\rho=-p$ (see \cite{FodPer}) 
we only deal here with the more general case (\ref{petcond1}).
Considering the other possible algebraically special types, the Petrov
type cannot be III because of $\Psi_0=\Psi_4=0$, and the Petrov II and
N cases also have the non-physical equation of state
$\rho=-p$. Finally, due to a theorem by Collinson \cite{Collinson},
the conformally flat case is also excluded.
 
Hence, the only algebraic special solutions of physical interest one might
hope to find are of Petrov type D.
Up to second order in the rotational parameter equation (\ref{petcond1})
gives only one real condition.
By substitution of the zeroth and first order field equations
(\ref{G110}) and (\ref{G30}) into (\ref{petcond1}) the {\it Petrov
type D} condition gives the relation \cite{Fodor}
\begin{equation}\label{Petrov}
\left(rB\frac{dB}{dr}+1-B^2\right)(h_2-m_2)
=\frac{r^4 A^2}{6}\left[\frac{d}{dr}
\left(\frac{B^2\omega}{A^2}\right)\right]^2.
\end{equation}

\subsection{Field equations for the interior region}

Note that $m_0$ and $h_0$ do not appear in equations (\ref{G110}),
(\ref{G30}), (\ref{G12}), (\ref{G22}), (\ref{G1122}) and
(\ref{Petrov}) and hence this subsystem for $A$, $B$, $\omega$, $m_2$,
$k_2$ and $h_2$ decouples. Notice that these equations contain $m_2$
only algebraically. In section \ref{Numericalsolutions} the system
will be reformulated as a coupled system of six first order ordinary
differential equations.  Due to the requirement of a regular centre
the solutions to this subsystem will only depend on four constants of
integration.

\subsection{Vacuum metric}\label{sectionvacuum}

In the exterior vacuum region we will use a frame adapted to the
asymptotically non-rotating observer.  Solving the field equations
detailed in Section \ref{sectionfieldeq} by imposing $p=\rho=0$ the
metric functions for the vacuum region are given as follows
\cite{Hartle, bfmp}
\begin{equation}\label{eq0vacuum}
A^2=B^2=1-2M/r
\end{equation}
\begin{equation}\label{eq1vacuum}
\omega =\frac{2aM}{r^3}\ ,
\end{equation}
\begin{eqnarray}\nonumber
h_0 &=&\frac 1{r-2M}\left( \frac{a^2M^2}{r^3}+\frac r{2M}c_2\right) \\\nonumber
m_0 &=&\frac 1{2M-r}\left( \frac{a^2M^2}{r^3}+c_2\right) \\\nonumber
h_2 &=&3c_1r\left( 2M-r\right) \log \left( 1-\frac{2M}r\right) +a^2\frac
M{r^4}\left( M+r\right)   \\\nonumber
&&+2c_1\frac Mr\left( 3r^2-6Mr-2M^2\right) 
\frac{r-M}{2M-r}\\\nonumber
&&+\left( 1-\frac{2M}r\right) 
r^2q_1  \label{eqh2} \\\nonumber
k_2 &=&3c_1(r^2-2M^2)\log \left( 1-\frac{2M}r\right) -a^2\frac M{r^4}(2M+r) 
 \\\nonumber
&&-2c_1\frac Mr(2M^2-3Mr-3r^2)+\left( 2M^2-r^2\right) q_1 \\
m_2 &=&6a^2\frac{M^2}{r^4}-h_2 \ .  \label{eqk2}
\end{eqnarray}
In this approximation, the slowly rotating solution is characterized
by the mass $M$, the first order small rotation parameter $a$, and the
second order small constants $c_1$, $c_2$ and $q_1$.  When $q_1$ takes
the value zero the metric is known to be the general asymptotically
flat stationary and axisymmetric vacuum metric to second order (see
e.g. \cite{bfp}). It can be checked by plugging the vacuum quantities
into the Petrov type D condition (\ref{Petrov}) that the solution is
of Petrov type D only if both $c_1$ and $q_1$ are zero. The metric is
then equivalent to the Kerr metric to second order with mass $M
\rightarrow M-c_2$.

When $q_1 \neq 0$ the metric cannot be asymptotically flat.  It is
important to keep in mind, however, that without the inclusion of this
constant the matching conditions on the zero pressure surface are
overdetermined in general \cite{MarsSenovilla,bfmp}.

\section{Matching}\label{Matching}

The matching of the fluid ball to a suitable exterior vacuum region
happens at the zero pressure surface. Before matching these two
spacetime regions it is informative to investigate first the structure
of the constant pressure surfaces.

\subsection{The constant pressure surfaces}\label{constp}

The pressure in the rotating fluid configuration is given by the
function $p(r,\theta)=p_0(r)+p_2(r,\theta)$.  The surfaces of constant
pressure, ${\mathcal S}_{\bar r}$, may be labelled by the function
$\bar r$ defined by the relation
\begin{equation}
p(r,\theta)=\bar p_0(\bar r)\equiv p_0(\bar r)+\delta p(\bar r)\ ,
\end{equation}
where $p_0$ is the corresponding pressure for the non-rotating
configuration and $\delta p(\bar r)$ is a second order small shift of
the pressure that changes monotonously from the centre where it takes
the value $p_{20}(0)$ to the zero pressure surface where it becomes
zero.  The value of the central pressure to second order follows by
assuming regularity at the centre and by making use of the field
equations (\ref{G1120}) and (\ref{uncheos}), along with the relation
for $p_{20}$ (\ref{p20}) (see section \ref{sec:series}).  It turns out
that it will depend, among others, on one freely specifiable constant
corresponding to a spherically symmetric perturbation that produces a
second order small change of the central pressure (and density).  If
we choose to consider only rotational perturbations with
$p_{20}(0)=0$, $\delta p(\bar r)$ may be chosen to be identically
zero.

The radial displacement $\xi$ is defined by
\begin{displaymath}
r=\bar r+\xi\ .
\end{displaymath}
To second order one has
\begin{eqnarray}\nonumber
p(r,\theta)&=&p_0(r)+p_2(r,\theta)\\\nonumber
&=&p_0(r)+p_{20}(r)+p_{22}(r)P_2(\cos\theta)\\\nonumber
&=&p_0(\bar r)+\xi \left.\frac{dp_0}{dr}\right\vert_{\bar r}
+p_{20}(\bar r)+p_{22}(\bar r)P_2(\cos\theta)\\\nonumber
&\equiv& p_0(\bar r)
+\delta p(\bar r)\ ,
\end{eqnarray}
implying that $\xi$ possesses the form
$\xi=\xi_0+\xi_2P_2(\cos\theta)$, where $\xi_0$ and $\xi_2$ are given
as
\begin{eqnarray}\nonumber 
\xi_0=-\left[p_{20}(\bar r)-\delta p(\bar r)\right]
/(dp_0/dr\vert_{\bar r}) 
\end{eqnarray}
and
\begin{eqnarray}\label{xi} 
\xi_2=-p_{22}(\bar r)/(dp_0/dr\vert_{\bar r})\ .
\end{eqnarray}
Note that there will be a certain arbitrariness in $\xi_0$, the
average shift of the radius, unless $\delta p(\bar r)$ is
specified. At the origin $\xi_0=0$ and on the zero pressure surface
$r_1$ the expressions (\ref{pp}) and (\ref{p20}) together with $\delta
p(r_1)=0$ gives
\begin{eqnarray}\nonumber
\xi_0&=&
\frac{1}{12r B\frac{dA}{dr}\frac{d}{dr}\left(\frac{A}{B}\right)}\times
\\\label{xi0}
&&\left.\left[12r A^2\frac{dh_0}{dr}-12m_0\frac{d}{dr}\left(rA^2\right)
+r^4\left(\frac{d\omega}{dr}\right)^2\right]\right
\vert_{r=r_1}
\end{eqnarray}
for $\xi_0$. If we choose $\delta p(\bar r)\equiv 0$ this expression,
with $r_1$ substituted with $\bar r$, holds for any $\bar r$ in the
interval $[0,r_1]$.  From (\ref{pp}) and (\ref{p22}) $\xi_2(\bar r)$
is given by
\begin{equation}\label{xi2}
\xi_2= -\left.\frac{\left(3A^2h_2+ r^2\omega^2\right)}
{3A\frac{d A}{d r}}\right\vert_{r=\bar r} \, .
\end{equation}

The circumference of the intersection of a constant pressure surfaces
${\mathcal S}_{\bar r}$ and the equatorial plane $\theta=\pi/2$, which
is in fact a circle, is obtained from
\begin{eqnarray}\nonumber
d l^2 &=& (1+2k)r^2\sin^2\theta d\varphi^2 \\\nonumber
&=&(1+2k_0-k_2)
\left(\bar r+\xi_0-\frac{\xi_2}{2}\right)^2 d\varphi^2
\end{eqnarray}
giving
\begin{displaymath}
l_1=2\pi \bar r\left(1+k_0+\frac{\xi_0}{\bar r}-\frac{k_2}{2}
-\frac{\xi_2}{2\bar r}\right).
\end{displaymath}
The length of the curve $\gamma$, yielded by the intersection of ${\mathcal
S}_{\bar r}$ and a plane including the axis of rotational symmetry
is given as
\begin{displaymath}
l_2=2\pi \bar r\left(1+k_0+\frac{\xi_0}{\bar r}+\frac{k_2}{4}
+\frac{\xi_2}{4\bar r}\right),
\end{displaymath}
where we have used the relation 
\begin{eqnarray}\nonumber
&&d l^2 = (1+2k)r^2 d\theta^2=\\\nonumber
&&(1+2k_0+2k_2P_2(\cos\theta))
\left(\bar r+\xi_0+\xi_2 P_2(\cos\theta)\right)^2 d\theta^2,
\end{eqnarray}
along with the fact that the term obtained by substituting
$dr=-3\xi_2\cos\theta\sin\theta d\theta$ into the line element
(\ref{ds}) is of fourth order and is hence dropped.  The constant
pressure surfaces are oblate iff $l_1>l_2$, i.e. whenever
\begin{equation}\label{oblate}
k_2+\frac{\xi_2}{\bar r}<0. 
\end{equation}

Another way of determining the oblateness of the constant pressure surfaces is possible
by comparing the radial distance from the origin to the curve 
$\gamma$ and the analogous distance to second order in the eccentricity 
parameter $\epsilon$, $r=a(1-\frac{1}{2}\epsilon^2\cos^2\theta)$, for 
an ellipse in $\mathbb{R}^2$ with semi-major axis $a$. The curve $\gamma$ is then found to be an ellipse up to second order with
$\epsilon^2$ given as
\begin{equation}\label{epsilon2}
\epsilon^2=\left.-3\left[\frac{\xi_2}{B} + \int_0^{\bar r} \frac{m_2}{B} d\bar r\right]
\right/\int_0^{\bar r} \frac{d\bar r}{B} \, .
\end{equation}
 For infinitesimally small values of $\bar r$ (\ref{epsilon2}) reduces to
\begin{displaymath}
\epsilon^2 = -3\left(m_2+\frac{\xi_2}{\bar r}\right)  \, ,
\end{displaymath}
according to which the constant pressure surfaces are oblate iff
\begin{equation}
m_2+\frac{\xi_2}{\bar r}<0.
\end{equation}
Notice that this inequality, along with the relations (\ref{serieh0}) 
and (\ref{serieh1}) in section \ref{sec:series}, does also justify that 
the two different characterization of oblateness are compatible.

Note, finally, that $d\bar r$ is a form field orthogonal to the
constant pressure surfaces ${\mathcal S}_{\bar r}$.  Up to second
order the corresponding normalized field is
\begin{equation}
n_a=(0,(1+m)/B,3\xi_2\sin\theta\cos\theta/B,0) \ .
\end{equation}

\subsection{The matching}

In this section we match the interior rotating fluid solution to an
exterior vacuum region at the zero pressure surface $\mathcal{S}_{\bar
r=r_1}$.

In the vacuum exterior region, suitable hypersurfaces for matching are
determined by the condition \cite{Roos}
\begin{equation}
\tilde \Omega ^{2}g_{\varphi \varphi }+2\tilde \Omega g_{\varphi t}
+g_{tt}=1-\tilde C
\label{roosc}
\end{equation}
where $\tilde \Omega $ and $\tilde C$ are constants.  To second order
such a surface can be given as
\begin{equation}
r=r_1+\chi=r_1+\chi_0+\chi_2 P_2(\cos\theta)\ ,
\end{equation}
where $\chi_0$ and $\chi_2$ are second order small constants.  The
unit normal to this surface is given by
\begin{equation}
n_a^{(v)}=(0,(1+m^{(v)})/B^{(v)},3\chi_2\sin\theta\cos\theta/B^{(v)},0),
\end{equation}
where the uppercase index $\ ^{(v)}$ here and after refers to vacuum
quantities.

To adjust the coordinates in the two regions we apply a rigid rotation
in the interior by the transformation $\varphi \rightarrow
\varphi+\Omega t$ . Also, we can rescale the interior time coordinate
by a constant $c_4$ while matching the spherical basis solutions, and
then later by a second order small constant when doing the matching of
the corresponding rotating configuration $t \rightarrow c_4(1+c_3)t$.
We do not have such a freedom in choosing the time coordinate and
applying rotation in the exterior region since we want a coordinate
system adapted to asymptotically non-rotating stationary observers.

Together with the values of the other parameters, the location of the
matching surface ${\cal{S}}_{r_1}$ is determined by the Darmois-Israel
conditions \cite{Darmois,Israel}. In particular, these conditions
pick out the zero pressure surface as the matching surface.

The Darmois-Israel conditions require that the induced metrics agree
on the matching surface ${\cal{S}}_{r_1}$
\begin{equation}\label{indmetric}
ds^2\vert_{{\cal{S}}_{r_1}}=ds^2_{(v)}\vert_{{\cal{S}}_{r_1}}\ ,
\end{equation}
as well as, the induced second fundamental forms
\begin{equation}\label{indsecond}
K\vert_{{\cal{S}}_{r_1}}=K^{(v)}\vert_{{\cal{S}}_{r_1}},
\end{equation}
where $K$ is defined as  
\begin{displaymath}
K\equiv K_{ab}dx^adx^b\equiv h_a^{\,\, c}h_b^{\,\, d}n_{(c;d)}dx^adx^b,
\end{displaymath}
with
\begin{displaymath}
{h_a}^b=n_an^b+{\delta_a}^b
\end{displaymath}
being the projection operator onto the hypersurface orthogonal to the
normal vector $n_a$.

Writing $g_{ab}$ and $K_{ab}$ as 
\begin{eqnarray}\nonumber
g_{ab}&=&g^{(0)}_{ab}+g^{(1)}_{ab}+g^{(2)}_{ab} \\\nonumber
K_{ab}&=&K^{(0)}_{ab}+K^{(1)}_{ab}+K^{(2)}_{ab}\, ,
\end{eqnarray}
where the superscripts $^{(0)},^{(1)},^{(2)}$ indicate zeroth, first
and second order terms respectively, one obtains to second order
\begin{eqnarray}\nonumber
g_{ab}(r)&=&g^{(0)}_{ab}(r)+g^{(1)}_{ab}(r)+g^{(2)}_{ab}(r)\\\nonumber
&=&g^{(0)}_{ab}(r_1+\xi)+g^{(1)}_{ab}(r_1)+g^{(2)}_{ab}(r_1)\\ \nonumber
&=&g^{(0)}_{ab}(r_1)+\left.\frac{\partial g^{(0)}_{ab}(r)}
{\partial r}\right\vert_{r=r_1}
\negthinspace\negthinspace\negthinspace\negthinspace\negthinspace
\negthinspace\negthinspace\negthinspace\xi
+g^{(1)}_{ab}(r_1)+g^{(2)}_{ab}(r_1)\, , \\ \label{gabexpand}
\end{eqnarray}
and similarly for $K_{ab}$, on the matching surface given by
$r=r_1+\xi$.  An analogous result holds in the outer region.  From
(\ref{indmetric}), (\ref{indsecond}) and (\ref{gabexpand}) we then
obtain the following equations to order by order.

Zeroth order:
\begin{equation}
A^{(v)}=c_4A\ ,\ \ \ 
B^{(v)}=B\ ,\ \ \ 
A^{(v)}_{\ \ ,r}=c_4A_{,r}
\end{equation}

 First order:
\begin{equation}
\omega^{(v)}=c_4(\omega-\Omega)\ ,\ \ \
\omega^{(v)}_{\ \ ,r}=c_4\omega_{ ,r} 
\end{equation}
  
Second order:
\begin{equation}
h_2^{(v)}=h_2\ ,\ \ \
k_2^{(v)}=k_2\ ,\ \ \
h_0^{(v)}=h_0+c_3
\end{equation}

\begin{equation}
c_4A(h_{0\ ,r}^{(v)}-h_{0 ,r})+\xi_0(A^{(v)}_{\ \ ,rr}-c_4A_{,rr})=0
\end{equation}

\begin{eqnarray}\nonumber
&&c_4A(h_{2\ ,r}^{(v)}-h_{2 ,r})+
\xi_2(A^{(v)}_{\ \ ,rr}-c_4A_{,rr})-\\
&&c_4rA_{,r}(k_{2\ ,r}^{(v)}-k_{2 ,r})=0
\end{eqnarray}

\begin{equation}
B(m^{(v)}_0-m_0)=\xi_0(B^{(v)}_{\ \ ,r}-B_{,r})
\end{equation}

\begin{equation}
B(m^{(v)}_2-m_2)=\xi_2(B^{(v)}_{\ \ ,r}-B_{,r})+rB(k_{2\ ,r}^{(v)}-k_{2 ,r})
\end{equation}

\begin{equation}
\chi_0=\xi_0\ , \ \ \ \chi_2=\xi_2 \ .
\end{equation}
All quantities here are evaluated at $r=r_1$, i.e. at the zeroth order
radius.  From the zeroth order equations we solve for $M$ and $c_4$ as
\begin{equation}
M=\frac{1}{2}r_1(1-B^2) \,\, , \,\, c_4=\frac{B}{A} \ .
\end{equation}
The third equation is equivalent to the zero pressure condition and
also gives the radius $r_1$ implicitly from
\begin{equation}
r_1=\frac{A}{2B^2\frac{dA}{dr}}(1-B^2) \, .
\end{equation}
To first order we solve for $a$ and $\Omega$
\begin{equation}
a=\frac{Br_1^3}{3A(B^2-1)}\frac{d\omega}{dr}\,\, , 
\,\, \Omega=\frac{r_1}{3}\frac{d\omega}{dr}+\omega \, .
\end{equation}
 From six of the nine second order equations we can solve for $c_1$,
$c_2$, $c_3$, $q_1$, $\chi_0$ and $\chi_2$ as
\begin{widetext}
\begin{equation}
c_1=\frac{B^2\left[r_1^4B^2(B^4-3)\left(\frac{d\omega}{dr}\right)^2+
36A^2h_2(1-B^4)+72A^2B^2(h_2+k_2)\right]}{9(B^2-1)^6A^2r_1^2}
\end{equation}
\end{widetext}
\begin{equation}
c_2=\frac{\xi_0}{2}\left(B^2-1+2r_1B\frac{dB}{dr}\right)
-r_1B^2m_0 -
\frac{r_1^5}{36}\frac{B^2}{A^2}\left(\frac{d\omega}{dr}\right)^2
\end{equation}
\begin{equation}
c_3=\frac{r_1^4}{36A^2}\left(\frac{d\omega}{dr}\right)^2
+\frac{c_2}{r_1B^2(1-B^2)}-h_0
\end{equation}
\begin{eqnarray}\nonumber
q_1&=&\frac{1}{9r_1^2A^2(B^2-1)^6}\left[18k_2A^2(B^4-1)(B^4-8B^2+1)
 \right.
\\\nonumber
&& +216A^2B^2\ln B\left(2B^2(h_2+k_2)+  
 (1-B^4)h_2\right)+\\\nonumber 
&&36h_2A^2B^2(B^2-1)(B^4+B^2-8)+ \\\nonumber
&&r_1^4\left(\frac{d\omega}{dr}\right)^2B^2\left((B^2-1)(2+11B^2-7B^4)+ \right.
\\
&& \left. \left.
6B^2(B^4-3)\ln B \right)\right]
\end{eqnarray}
\begin{equation}
\chi_0=\xi_0\quad\hbox{and}\quad \chi_2=\xi_2\ ,
\end{equation}
where $\xi_0$ and $\xi_2$ are obtained from (\ref{xi0}) and
(\ref{xi2}).  The remaining three equations turn out to be identically
satisfied due to the other matching equations and the field
equations. Note that we did {\it{not}} assume the Petrov type D
condition (\ref{Petrov}) when calculating the matching conditions.
Hence an appropriate matching can be done, i.e. the vacuum metric in
section \ref{sectionvacuum} is general enough for describing the
exterior of any axisymmetric rigidly rotating perfect fluid ball up to
second order.

\section{Numerical integration}\label{Newvariables}

In this section we provide a reformulation of the field equations
which is more suitable for numerical integration.  By doing this we
can get higher precision at the origin where apparent singularities
arise, moreover the freely specifiable constants are identified more
easily this way.

\subsection{Integrating the zeroth order field equation}

In order to simplify (\ref{G110}) it is convenient to redefine the
functions $A$, $h$, $m$ and $k$ in terms of the function $\nu$ as
\begin{equation}
A=e^{\nu}, \quad h=\tilde h e^{-2\nu}, \quad m=\tilde m e^{-2\nu}, 
\quad k=\tilde k e^{-2\nu}
\end{equation}
giving
\begin{eqnarray}
ds^{2}  &=&e^{2\nu}(1+2\tilde he^{-2\nu})dt^{2}
-(1+2\tilde me^{-2\nu})\frac{1}{B^2}dr^{2}\nonumber\\
& &  -(1+2\tilde ke^{-2\nu})r^{2}\left[  
d\theta^{2}+\sin^{2}\theta\left(
d\varphi-\omega dt\right)  ^{2}\right]  \ . \label{ds2}\nonumber
\end{eqnarray}
The equations simplify considerably due to the fact that only the
derivative of $\nu$ will appear. Hence we introduce the function $z$
by
\begin{equation}
\frac{z}{B}=r\frac{d\nu}{dr}+1 \, .
\end{equation}
Then the zeroth order equation (\ref{G110}) becomes first order in $z$
and algebraic in $B$ \cite{Fodorsph}
\begin{equation}\label{zeq}
B r \frac{dz}{dr}+2B^2+z^2-4Bz+1=0 \ ,
\end{equation}
furthermore, the pressure of the non-rotating configuration (\ref{pp})
takes the form
\begin{equation}
p_0=\frac{1}{r^2}\left(2Bz-B^2-1\right) \ .
\end{equation}

\subsection{Series expansion around a regular centre}\label{sec:series}

 For sufficiently regular configurations close to the centre the metric
coefficients can be given as power series in $r$.  Assuming that the
central pressure and density are finite it follows that $B(0)=z(0)=1$.
The assumption of smoothness of the configurations at the symmetry
centre, in the spacetime sense, implies that the odd coefficients in
the expansions of the basic variables are zero. Although the smoothness 
implies the vanishing of 
the odd coefficients without the use of the field equations, the 
requirement of smoothness of central density and pressure, together with 
the field equations, also implies the smoothness of the metric functions.

The vanishing of the odd coefficients can be shown in the generic case
by plugging power series expansions of the dependent variables into
the field equations. In these cases the smoothness of the density and
pressure results from these considerations.  If odd powers of $r$ are
included in the expansion of the metric variables then it can be shown
that the field equations and the Petrov D condition, together with the
assumption of finite and positive central pressure and density, imply
the vanishing of the coefficients of all odd terms.  However, odd
terms may appear in special cases, and then the metric ceases to be
smooth at the origin.  For example, consider the spherically symmetric
case when the only field equations is given by (\ref{zeq}).  If $r^3$
terms are included in the expansion of $B$ then one also obtains a non
smooth density gradient $\frac{d \rho_0}{d r}\vert_{r=0}\neq
0$. However, since the field equation implies that the pressure
gradient $\frac{d p_0}{d r}\vert_{r=0}$ is always zero, the squared
speed of sound is vanishing at the origin, i.e. $v_s^2=\frac{d p_0}{d
\rho_0}\vert_{r=0}=0$.  It is of interest to find the minimal
requirements needed to guarantee that the solution is regular to any
order. A conjecture is that this is the case when the central pressure
and density are finite and $\frac{d p}{d \rho}\vert_{r=0}\neq0$
\cite{Fodorsmooth}. If the Petrov D condition is used instead of an
equation of state to specify the configuration then it can be shown
that $\frac{d p}{d \rho}\vert_{r=0}=0$ can occur only if either the
central pressure or density is negative.

Hence, assuming a smooth centre in the spacetime sense, the odd powers
will be omitted hereafter.  Plugging the expressions
\begin{eqnarray}\nonumber
B&=&1+b_1 r^2+b_2 r^4+...\\\nonumber
z&=&1+z_1 r^2+z_2 r^4+...\\\nonumber
\omega&=&\omega_0+\omega_1 r^2+\omega_2 r^4+...\\\nonumber
\tilde h_2&=&h_2^{(0)}+h_2^{(1)}r^2+h_2^{(2)}r^4...\\\nonumber
\tilde m_2&=&m_2^{(0)}+m_2^{(1)}r^2+m_2^{(2)}r^4...\\\label{expansion1}
\tilde k_2&=&k_2^{(0)}+k_2^{(1)}r^2+k_2^{(2)}r^4...
\end{eqnarray}
into the field equations justifies then that all coefficients can be
given in terms of $b_1$, $z_1$, $\omega_0$ and $h_1\equiv h_2^{(1)}$.
To zeroth order one obtains
\begin{equation}\label{serieh0}
h_2^{(0)}=m_2^{(0)}=k_2^{(0)}=0 \, ,
\end{equation}
then to second order
\begin{equation}\label{serieh1}
m_2^{(1)}=k_2^{(1)}=-h_2^{(1)}\equiv -h_1
\, , \,\,\, \omega_1=\frac{2}{5}\omega_0(z_1-3b_1)\, ,
\end{equation}
and finally to fourth order
\begin{eqnarray}\nonumber
b_2&=&-\frac{b_1^2}{2}+\frac{3\omega_0^2}{50 h_1}
\left(z_1-3b_1\right)^2, \;\;\; 
z_2=b_1z_1-\frac{z_1^2}{2}-b_1^2\\\nonumber
h_2^{(2)}&=&\frac{h_1(3z_1-13b_1)}{14}-\\\nonumber
&&\frac{\omega_0^2(z_1-3b_1)
\left[
2 h_1 (22z_1-31b_1)+3\omega_0^2(z_1-3b_1)\right]}
{210 h_1 (z_1-b_1)}\\\nonumber
k_2^{(2)}&=&\frac{\omega_0^2}{6}(z_1-3b_1)
+\frac{h_1}{2}(b_1-z_1)-h_2^{(2)},\\\nonumber 
m_2^{(2)}&=&\frac{2\omega_0^2}{3}(z_1-3b_1)-h_2^{(2)}\\\nonumber
\omega_2&=&\frac{\omega_0(z_1-3b_1)}{70 h_1}
\left[h_1(z_1-33b_1)-3\omega_0^2(z_1-3b_1)\right] .\\
\end{eqnarray}

The expansion of the density and pressure of the nonrotating
configuration can be written as
\begin{eqnarray}\label{densexp}
\rho_0&=&\rho_{0c}-\frac{3\omega_0^2}{20h_1}
\left(p_{0c}+\rho_{0c}\right)^2r^2+O(r^4) \\
p_0&=&p_{0c}-\frac{1}{12}\left(3p_{0c}^2+4\rho_{0c}p_{0c}
+\rho_{0c}^2\right)r^2+O(r^4)
\end{eqnarray}
where the central density and pressure are given by
\begin{equation}
\rho_{0c}=-6b_1\ ,\ \ \ p_{0c}=2z_1 \ .
\end{equation}
This shows that for realistic configurations $b_1<0$ and $z_1>0$,
consequently the $z_1-b_1$ term in the denominator of $h_2^{(2)}$ is
nonvanishing. Also, the existence of a local maximum of the density at
the center implies $h_1>0$. The pressure always has a local maximum at
$r=0$ if the central values are positive.

Assuming $h_1=0$ implies that the higher coefficients in the expansion
of $h_2$, $m_2$ and $k_2$ are zero.  We conjecture that
$h_2=m_2=k_2\equiv0$, which is also supported by a numerical
calculation.  From this it follows that $\omega=\omega_0=$constant and
that $A=B=\sqrt{1+Cr^2}$. But this simply gives the de Sitter or anti
de Sitter solutions, depending on the sign of the integration constant
$C$, in a rotating frame.

Plugging the expansions of the original (no tilde) second order
spherical perturbation quantities
\begin{eqnarray}\nonumber
h_0 &=&h_0^{(0)}+h_0^{(1)}r^2+h_0^{(2)}r^4+...\\\label{expansionh0}
m_0&=&m_0^{(0)}+m_0^{(1)}r^2+m_0^{(2)}r^4+...
\end{eqnarray}
together with (\ref{expansion1}) into the two remaining equations
(\ref{G1120}) and (\ref{uncheos}) gives that two constants,
e.g. $h_0^{(0)}$ and $m_0^{(1)}$ are freely specifiable, whereas the
other coefficients can be expressed in terms of these two. Of these
only $m_0^{(1)}$ is essential since the constant $h_0^{(0)}$ can be
absorbed by a second order rescaling of the time coordinate. To second
order one gets
\begin{equation}
m_0^{(0)}=0 \quad \hbox{and} \quad 
h_0^{(1)}=\frac{5m_0^{(1)}h_1(z_1-b_1)}{2\omega_0^2(z_1-3b_1)}
+\frac{m_0^{(1)}}{2}
\end{equation}
unless $z_1=3b_1$ corresponding to $p_{0c}=-\rho_{0c}$.  From this
result it follows that $p_{20}$, as given by (\ref{p20}), at the
origin is
\begin{equation}\label{p200}
p_{20}(r=0)=\frac{10 m_0^{(1)} h_1\left(b_1-z_1\right)}
{\omega_0^2\left(3 b_1-z_1\right)} \ .
\end{equation}
Hence the central pressure will be unchanged if $m_0^{(1)}=0$. However,
the higher order coefficients in the expansions
(\ref{expansionh0}) are still nonzero, i.e.\ the expansions of $h_0$
and $m_0$ starts with $r^4$ terms. We note that, in general,
$\rho_{20}(r=0)=6m_0^{(1)}$ and $p_{20}(r=0)=4h_0^{(1)}-2m_0^{(1)}$.

Purely spherically symmetric perturbations, corresponding to a small
change of central pressure but unchanged equation of state, of the
type (\ref{expansionh0}) can be obtained. This kind of perturbations are
possible even when there is no rotation at all. By choosing
$\omega_0=h_1=0$ and $m_0^{(1)}\neq 0$ a spherically symmetric
perturbation, with a second order shift of the central pressure given
by
\begin{equation}
p_{20}(r=0)=\frac{6 m_0^{(1)} \left(b_1-z_1\right)\left(3 b_1-z_1\right)}
{5 b_1^2 + 10 b_2} 
\end{equation}
is produced. This expression remains valid in general, even when the
Petrov D condition is not assumed.

\subsection{System of differential equations}

Motivated by the results of the previous section it is advantageous to
define the new dependent variables $\beta$, $\zeta$, $\tilde y$, $\hat
h$, $\hat k$ and $\hat m$ through
\begin{eqnarray}\nonumber
B&=&1+r^2\beta\\\nonumber
z&=&1+r^2\zeta\\
\omega_{,r}&=&2r\tilde y\\\nonumber
\tilde h_2&=&r^2 \hat h\\\nonumber  
\tilde k_2&=&r^2(r^2\hat k-\hat h)\\\nonumber
\tilde m_2&=&r^2(r^2\hat m-\hat h)\, .
\end{eqnarray}

The closed subsystem of equations (\ref{G110}), (\ref{G30}), (\ref{G12}), (\ref{G22}), (\ref{G1122}) 
and (\ref{Petrov})  then takes the form
\begin{eqnarray}\nonumber
\frac{d\zeta}{dr}&=&-\frac{r\left(2\beta\left(\beta-\zeta\right)+\zeta^2\right)}{\beta r^2+1}\\\nonumber
\frac{d\beta}{dr}&=&\frac{2\omega^2\left(\zeta-3\beta\right)-
3\hat m}{2\omega^2 r\left(\beta r^2+1\right)}+\frac{\tilde y^2 r}{\omega^2}\left(\beta r^2+1\right)
+\frac{r\beta\left(\zeta-3\beta\right)}{\left(\beta r^2+1\right)}\\\nonumber
\frac{d\omega}{dr}&=&2\tilde y r
\end{eqnarray}
\begin{eqnarray}\nonumber
\frac{d\tilde y}{dr}&=&\frac{3\hat m -5\omega\tilde y}{r\omega\left(\beta r^2+1\right)^2}
-\frac{r\tilde y^2 }{\omega^2}\left(2\omega+\tilde y r^2\right)+\\\nonumber
&&\frac{r\tilde y\left(3\hat m -10\beta\omega^2(2+\beta r^2)\right)}{2\omega^2\left(\beta r^2+
1\right)^2}\\\nonumber
\frac{d\hat h}{dr}&=&\frac{\left[\hat m (3\hat h +\omega^2)-4\omega^2(\zeta\hat h+ \hat k 
-2\beta\hat h)\right]}
{2r\omega^2(\zeta-\beta)(\beta r^2+1)}\\\nonumber
&&-\frac{r\tilde y^2\left(2\omega^2+3\hat h\right)\left(\beta r^2+1\right)}{3\omega^2\left(\zeta-\beta\right)}\\
\nonumber
&&+\frac{r\left[r^2\hat m (\zeta-\beta)^2-\beta\hat h (2\zeta - 3\beta)\right]}
{(\zeta-\beta)(\beta r^2+1)}\\\nonumber
 \frac{d\hat k}{dr}&=&\frac{\hat m-4\hat k +2\hat h\left(\beta-\zeta\right)}
{r\left(\beta r^2+1\right)}+\frac{r\left(2\hat k \left(\zeta-3\beta\right)+\hat m\zeta\right)}{\beta r^2+1} \ , \\\label{system}
\end{eqnarray}
while $\hat m$ can be solved for algebraically as
\begin{widetext}
\begin{eqnarray}\nonumber
\hat m &=&\frac{2}{3}\omega^2(\zeta-3\beta)+\frac{2}{3}r^2\tilde y^2 (\beta r^2+1)^2+
\frac{2r^2\omega^2}{3}\left[\frac{\tilde y (\beta r^2+1)^2\left[2\omega (\zeta - 3\beta)
-\tilde y \left(1+\beta r^4 (\zeta-\beta)+2\beta r^2\right)\right]
 }
{r^2(\beta r^2+1)\left[\omega^2(\zeta-2\beta)-\tilde y (\beta r^2+1)(r^2\tilde y+2\omega)\right]+
3\hat h-\beta r^2\omega^2}\right]
\\\nonumber
&&+\frac{2r^2\omega^2}{3}\left[\frac{\beta (\zeta-2\beta)\left(3 \hat h - r^2 \omega^2 (\zeta-3\beta)\right)  -\omega^2 (\zeta -3 \beta)^2 }
{r^2(\beta r^2+1)\left[\omega^2(\zeta-2\beta)-\tilde y (\beta r^2+1)(r^2\tilde y+2\omega)\right]+
3\hat h-\beta r^2\omega^2}\right].
\end{eqnarray}
\end{widetext}
Boundary conditions at $r=0$ are given as
\begin{eqnarray}\nonumber
\beta(0)&=&b_1, \; \zeta(0)=z_1, \; \omega(0)=\omega_0, \; 
\\\nonumber
\tilde y(0)&=&\omega_1=\frac{2}{5}\omega_0(z_1-3b_1),
\hat h(0)=h_1, \\\nonumber
\; \hat k(0)&=&k_2^{(2)}+ h_2^{(2)}=\frac{\omega_0^2}{6}\left(z_1-3b_1\right)
+\frac{h_1}{2}\left(b_1-z_1\right) \ .
\end{eqnarray}
The relation
\begin{displaymath}
 \hat m(0)=m_2^{(2)}+h_2^{(2)}=\frac{2}{3}\omega_0^2(z_1-3b_1)
\end{displaymath}
is then satisfied identically.  As we have seen there are four freely
specifiable constants: $b_1$, $z_1$, $\omega_0$ and $h_1$.

The system of equations (\ref{system}) possesses two types of scale
invariances. The first one is associated with the rescaling of the
$r$-coordinate, $r \rightarrow \alpha r$, under which transformation
the dependent variables scale as
\begin{eqnarray}\label{scale1}
\;\;\beta,\zeta,\hat m,\hat k,\tilde y \rightarrow  \frac{\beta}{\alpha^2}, 
 \frac{\zeta}{\alpha^2}, 
\frac{\hat m}{\alpha^2}, \frac{\hat k}{\alpha^2},\frac{\tilde y}{\alpha^2}, \;\; \omega,\hat h \rightarrow \omega,\hat h \, .
\end{eqnarray} 
There is also a rescaling associated to the rescaling of the
rotational parameter $\omega_0$, following the rule $\omega_0
\rightarrow \gamma \omega_0$, which induces the transformation
\begin{eqnarray}\nonumber
&&\omega,\tilde y\rightarrow\gamma\omega,\gamma\tilde y, \;\; \hat h,\hat m,\hat k
\rightarrow\gamma^2\hat h,\gamma^2\hat m,\gamma^2\hat k, \\\label{scale2}
&&\beta, \zeta, r \rightarrow \beta, \zeta, r  \ .
\end{eqnarray}
It is interesting that a combination of the above two rescalings with
$\gamma=1/\alpha$ yields a similarity transformation of the
investigated system.

Due to these scale invariances of the equations two of the constants,
e.g., $\omega_0$ and $b_1$, can be fixed. All other configurations can
be obtained by rescaling.  Note also that $b_1$ and $z_1$ can be
expressed in terms of the zeroth order central density and pressure as
\begin{displaymath}
b_1=-\frac{1}{6}\rho_{0c}, \; z_1=\frac{1}{2}p_{0c}.
\end{displaymath}

Some of the above equations contain terms of the type `` $F/r$'',
thereby they are apparently singular at the origin, so we collected
them to the beginning of the right hand sides.  It can be checked case
by case that all of the corresponding numerators vanish at the
origin. Nevertheless, in determining the values of the corresponding
ratios numerically it turned out to be advantageous to use as the
fundamental variables the differences $\zeta_{\Delta}, \beta_{\Delta},
\omega_{\Delta}, \tilde y_{\Delta}, \hat h_{\Delta}, \hat m_{\Delta},
\hat k_{\Delta}$ between the variables $\zeta, \beta, \omega, \tilde
y, \hat h, \hat m, \hat k$ and their exact values $\zeta_0, \beta_0,
\omega_0, \tilde y_0, \hat h_0, \hat m_0, \hat k_0$ at the origin. Due
to the cancellation of the terms involving the exact values at the
origin what remains from the numerators will be proportional to $r^2$.

\section{Numerical solutions}\label{Numericalsolutions}

The system (\ref{system}), when rewritten in terms of the variables
$\zeta_{\Delta}, \beta_{\Delta}, \omega_{\Delta}, \tilde y_{\Delta},
\hat h_{\Delta}$, $\hat m_{\Delta}$, $\hat k_{\Delta}$, was solved
using fourth order Runge-Kutta. A check of the convergence factor
\begin{displaymath}
C_n=\frac{f_n-f_{2n}}{f_{2n}-f_{4n}} \, ,
\end{displaymath}
where $n$ is the number of points in a given $r$-interval, was
performed for various quantities $f$.  We found that the errors
decreased according to the expectations, i.e. the value of $C_n$ was
found to be close to $16$.

We also checked whether the scale invariance properties of the field
equations were reproduced properly by our numerical code.  When we
rescaled the freely specifiable boundary data according to
(\ref{scale1}) or (\ref{scale2}) then all the dependent variables
scaled in the appropriate way.

In scanning the four dimensional parameter space, due to the scaling
invariances, without loss of generality, we fixed $b_1=-1$ and
$\omega_0=0.1$ while we varied the central values $z_1$ and
$h_1$. Notice that to have positive central densities and pressures
the relations $b_1<0$ and $z_1>0$ also have to be satisfied. The
integrations were carried out until the zero pressure surface was
reached.

\subsection{Check of Wahlquist}\label{sec:Wahlquist}

The code was checked for the Wahlquist solution \cite{Wahl}, which is
of Petrov type D. To second order it is given by \cite{bfmp}
\begin{eqnarray} 
&&ds^{2} =f_{0}(1+2h)dt^{2}
-2\frac{1+2m}{\mu _{0}\kappa ^{2}f_{0}}dx^{2}- \nonumber \\ &&
\frac{2}{\mu _{0}\kappa ^{2}}\sin ^{2}x (1+2k)\left[d\theta ^{2}+\sin
^{2}\theta\left( d\varphi -\omega dt\right) ^{2}\right] \label{dsw}
\end{eqnarray}
with
\begin{equation}
f_{0}=1+\frac{1}{\kappa ^{2}}\left( 1-x\cot x\right) 
\end{equation}
and
\begin{equation}\label{Womega}
\omega^{\ } =\frac{\mu _{0}r_{0}}{2\sin ^{2}x}\left( 1-x\cot x\right)  \ .
\end{equation}
The transformation to the Hartle variables used in (\ref{hartlemetric}) is given by
\begin{displaymath}
r=\sqrt{\frac{2}{\mu_0 \kappa^2}}\sin x\ .
\end{displaymath}
 For the functions $h$, $m$ and $k$ see \cite{bfmp}. 
 For the Wahlquist solution the four starting values are given by
\begin{displaymath}
b_1=\frac{\mu_0}{12}(1-3\kappa^2),\,\, z_1=\frac{\mu_0}{4}(1-\kappa^2),\,\, \omega_0=\frac{\mu_0 r_0}{6}
\end{displaymath}
and
\begin{equation}
 h_1=-\frac{\mu_0^2r_0^2\kappa^2}{60}
\end{equation}
in terms of the three integration constants $\mu_0$, $\kappa$ and $r_0$. Solving for $h_1$ gives
\begin{equation}
h_1=\frac{(z_1-3 b_1)\omega_0^2}{5(b_1-z_1)}\, .
\end{equation}
Note that $h_1<0$ for positive central density and pressure, implying together with equation (\ref{densexp})
the well known fact that the density of the Wahlquist solution has a minimum at the centre.

\begin{figure} [!h]
\epsfxsize=3.2in
\epsffile{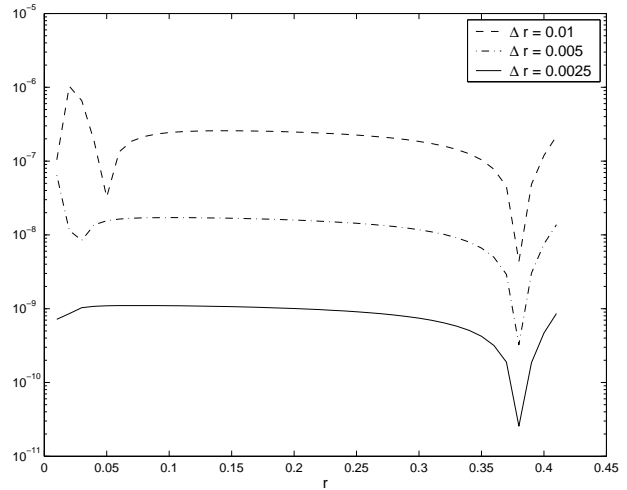} 
\vskip2mm
\caption{\small 
The relative error $|\omega_{\rm numerical}-\omega_{\rm analytic}|/\omega_{\rm analytic}$ 
is shown for the resolutions $\Delta r=0.01$,
$\Delta r=0.005$ and $\Delta r=0.0025$.
Starting values are $z_1=-b_1=1$ and $\omega_0=0.1$ giving $h_1=-0.004$.}
\hskip1.5cm
\label{fig:WWhaldif}
\end{figure}

In figure \ref{fig:WWhaldif} the relative error between the analytical solution for the
the rotational function $\omega$, as given by (\ref{Womega}),
and the numerical solution is plotted for various resolutions.

As it was already found \cite{bfmp,bfp} the shape of the Wahlquist
fluid ball is always prolate which is also in accordance with the
positive sign of the quantity $k_2+\xi_2/r_1$. The quantity $q_1$ is
not zero for the Wahlquist solution which is equivalent to that it
cannot be matched to an asymptotically flat exterior solution to
second order \cite{bfmp,bfp,SarnobatHoenselaers}.  The value of $c_1$
was also found to be negative for all the tested Wahlquist
configurations. This is also verified by analysing the analytical expression 
for $c_1$ given in \cite{bfp}.
It would be interesting to know whether the sign of this quantity is related to 
the shape in general.

\subsection{Asymptotically flat solutions}

\begin{figure} [!h]
\epsfxsize=3.5in
\epsffile{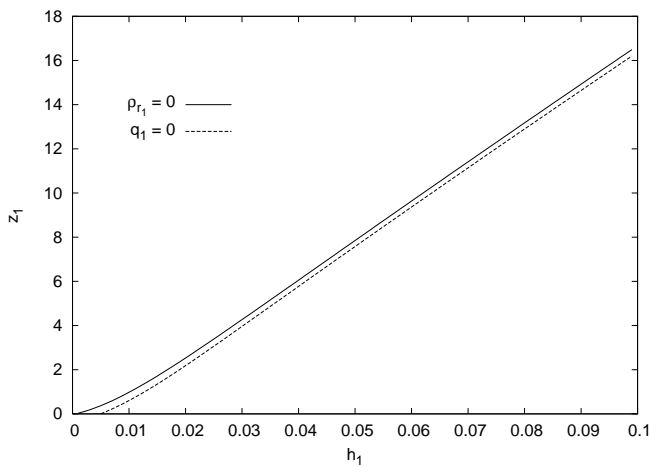} 
\vskip2mm
\caption{\small  Along the solid curve the density and pressure become
  zero for the same value of the radius $r=r_1$.  
Below this curve those configurations can be found which can be
matched to an exterior vacuum region.  
The dashed curve represents those configurations in the $z_1
h_1$-plane for which the exterior vacuum region is  
asymptotically flat.}
\label{fig:q1=0}
\end{figure}

\begin{figure} [!h]
\epsfxsize=3.4in
\epsffile{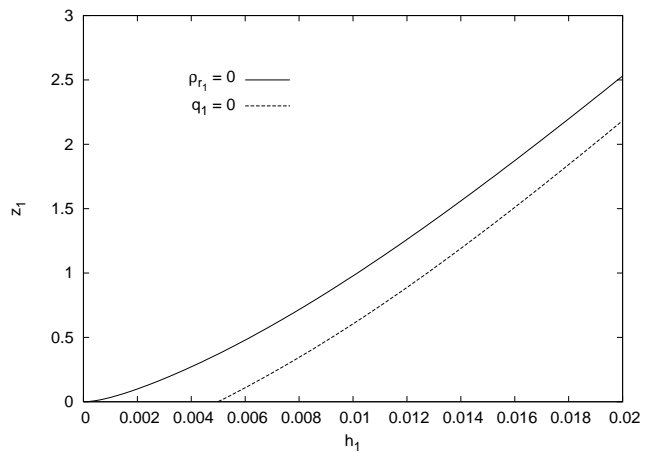} 
\vskip 2mm
\caption{\small The same curves as in Fig.\,\ref{fig:q1=0} are shown
  for small $z_1$, corresponding to low central pressures.} 
\label{fig:q1=00}
\end{figure}

A solution to the field equations (\ref{system}) is asymptotically
flat iff $q_1=0$. In Figure \ref{fig:q1=0} the points of the dashed
curve represent configurations with value $q_1=0$ in the
$z_1h_1$-plane, while in Figure \ref{fig:q1=00} a section of the same
curve for small $z_1$, corresponding to small central pressures, is
given.
Naturally, the asymptotically flat solutions, represented by points
belonging to these curves, always have finite radii, but further
increasing $z_1$, corresponding to the increase of the central
pressure, we get into a region where the density becomes negative
before a zero pressure surface is reached.  The limiting curve where
the pressure and density become zero at the same radius is shown in
 Figures \ref{fig:q1=0} and \ref{fig:q1=00} by the solid lines.  Due to
the negative density the pressure ceases to be a monotonic function of
$r$ for configurations above the $\rho_{r_1}=0$ curve.

Since for the asymptotically flat solutions $h_1$ may be seen as a
function of $z_1$, we see that $h_2$, $m_2$ and $k_2$ for these
solutions are determined by $b_1$, $z_1$ and $\omega_0$. The analysis
done in section \ref{sec:series} for the remaining field equations
showed that essentially one constant of integration is freely
specifiable for the functions $h_0$ and $m_0$. This implies then that
the second order configuration is completely determined by the zeroth
order spherical configuration, a second order spherically symmetric
perturbation and the magnitude of the rotation.

\subsection{Are there Kerr-like solutions?}

\begin{figure} [!h]
\epsfxsize=3.3in
\epsffile{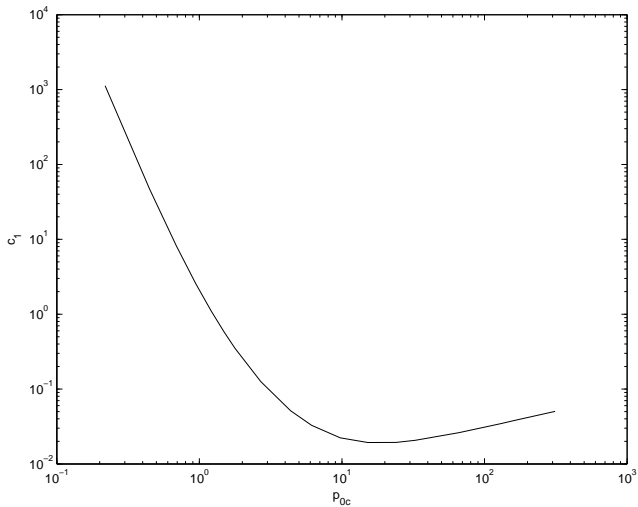} 
\vskip2mm
\caption{\small The constant $c_1$ as a function of zeroth order central 
pressure, $p_{0c}$, along the curve $q_1=0$.} 
\label{fig:c1min}
\end{figure}

To our knowledge the only known source for the Kerr metric is the thin
rotating disk of dust with $a=m$ found by Neugebauer and Meinel
\cite{NeugebauerMeinel}.  It is tempting to investigate whether there
can exist a fluid ball belonging to the class investigated in this
paper that could be matched to the Kerr solution to second order. To
settle this issue note first that the metric of the exterior region
becomes the Kerr metric with mass parameter $M-c_2$ iff $q_1=c_1=0$.
However, the numerical runs indicated that either $c_1>0$ for $h_1>0$
or $c_1<0$ for $h_1<0$ occurs, in general, i.e. the desired matching
seems not to be supported.  Note that these numerical findings are in
accordance with some earlier results, see
e.g. \cite{Bertietal,Hernandez}, telling that typically the exterior
metric deviates from the Kerr metric due to the ellipsoidal shape of
the rotating fluid ball.  To this end it is illuminating to consider
an expansion of the exterior metric for large $r$ which gives the
following leading terms of $g_{00}$ (with $q_1=0$)
\begin{equation}
g_{00}=1-\frac{2M\left(1-\frac{c_2}{M}\right)}{r}+
\frac{2MP_2(\cos\theta)\left(a^2+ \frac{16}{5}M^4c_1\right)}
{r^3}\ ,
\end{equation}
i.e., the associated quadrupole moment reads as (cf.,
e.g. \cite{Landau})
\begin{displaymath}
Q_{11}=Q_{22}=-Q_{33}/2=-2M\left(a^2+\frac{16}{5}M^4c_1\right)
\end{displaymath}
in an asymptotically Cartesian system with the 3-axis along the axis
of rotation.                                                       
In Figure \ref{fig:c1min} the value of $c_1$ as function of the
central pressure ${p_o}_c=2z_1$ along the $q_1=0$ curve is shown.

\subsection{Some asymptotically flat solutions with reasonable equation of state}

In this subsection we present some properties, like equation of state
and speed of sound, for some of the physically interesting inner fluid
ball configurations which can be matched to a suitable asymptotically
flat exterior vacuum region up to second order, i.e. those solutions
for which $q_1$ vanishes.  According to Table\,\ref{table:data} the
value of $h_1$ has to be smaller than around $0.012$ for these
solutions to have subluminal speed of sound, $v_s^2=d p/d \rho<1$. For
all configurations in this interval the speed of sound also increases
when approaching the centre, i.e. the fluid becomes stiffer as would
be expected on physical grounds.

\begin{figure} [!h]
\vskip1.0cm
\epsfxsize=3.3in
\epsffile{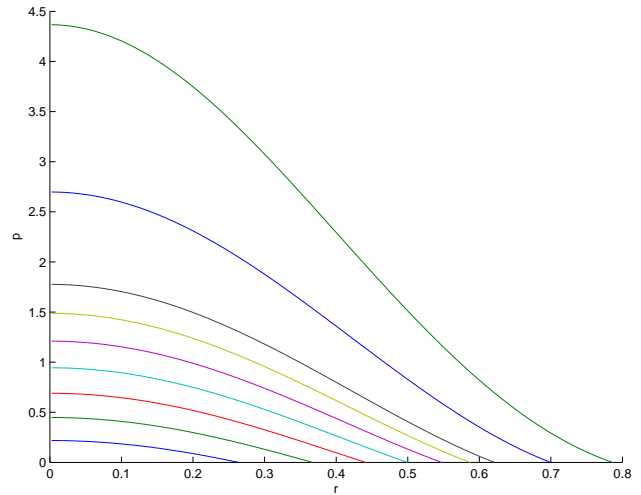} 
\vskip2mm
\caption{\small The zeroth order pressure $p_0$ is shown as function
  of $r$ for various central pressures along the curve $q_1=0$.  From
  top to the bottom the central pressures are given by $4.3666$,
  $2.6978$, $1.7762$, $1.4872$, $1.21$, $0.944$, $0.6896$, $0.448$ and
  $0.2184$ respectively, whereas the central density is
  $\rho_{0c}=6=-6b_1$.}
\label{fig:pressure}
\end{figure}

\begin{figure} [!h]
\epsfxsize=3.4in
\epsffile{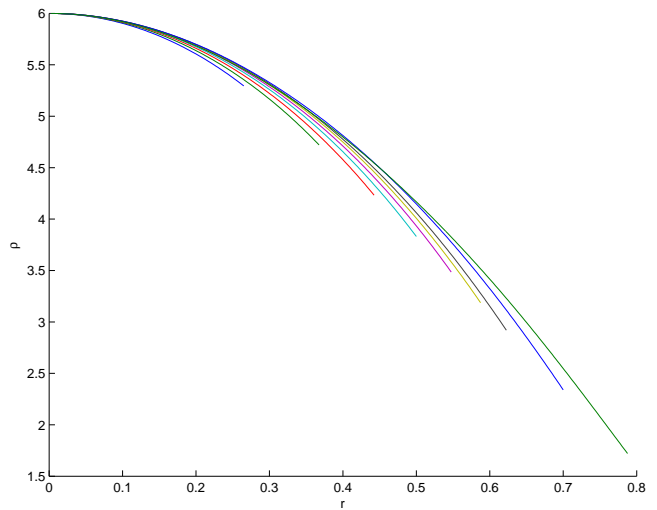} 
\vskip2mm
\caption{\small The zeroth order density $\rho_0$ is shown as function of $r$
  for various central pressures along the curve $q_1=0$.  
 From top to the bottom the central pressures are as given in figure \ref{fig:pressure}.}
\label{fig:density}
\end{figure}

\begin{figure} [!h]
\epsfxsize=3.4in
\epsffile{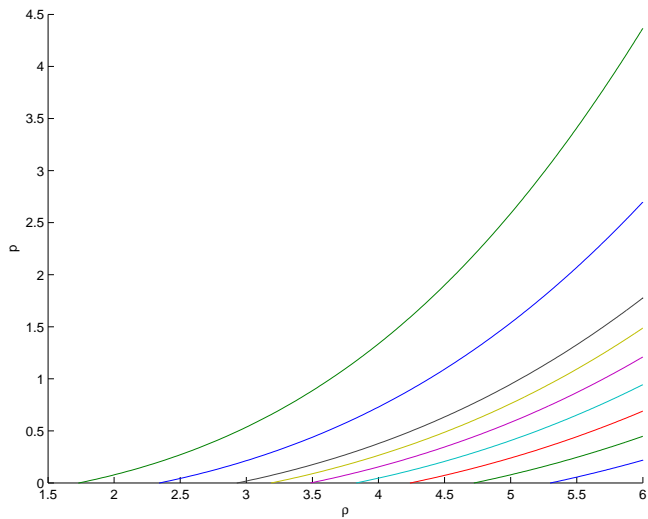} 
\vskip2mm
\caption{\small The equation of state $p=p(\rho)$ is shown for the
  same configurations as before.} 
\label{fig:eqstate}
\end{figure}

\begin{figure} [!h]
\epsfxsize=3.4in
\epsffile{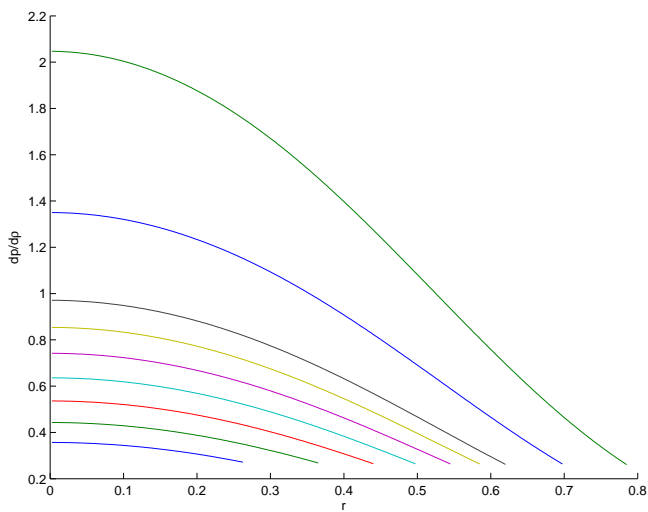} 
\vskip2mm
\caption{\small The square of the speed of sound, $v^2_s=\frac{d p}{d
    \rho}$, is shown for the same configurations as on the previous
  figures.} 
\label{fig:sound}
\end{figure}

In Figures \ref{fig:pressure} and \ref{fig:density} the zeroth order
pressure and density, $p_0$ and $\rho_0$, are shown as functions of
$r$ for some configurations with central pressure $p_{0c}$ between
$0.218$ and $4.366$, while in Figure \ref{fig:eqstate} and
\ref{fig:sound} the equation of state, i.e. $p$ as function of
$\rho$, and the square of the speed of sound, $v_s^2=\frac{d p}{d
\rho}$, as a function of $r$ are depicted, respectively, for the same
family of solutions.

Notice that the value of $v_s^2$ at the surface of the fluid ball
seems to be universal, i.e. apparently independent of the values of
the free parameters at the centre. Unfortunately, we could not find an
analytic argument supporting this observation, nevertheless, it is
interesting on its own right that such a lower limit may exist for the
class of the investigated fluid balls.

In general, the equations of state cannot be polytropic since the
density does not tend to zero while approaching the matching
surface. However, the equations of state can be approximated with a
polytropic one close to the centre. The two last columns in
Table\,\ref{table:data} provides the central values of the 
adiabatic index
\begin{equation}
\kappa \equiv \frac{n}{p}\frac{d p}{d n} = \frac{p+\rho}{p}\frac{d p}{d \rho} \, ,
\end{equation} 
where $n$ is the baryon number density, and the Newtonian adiabatic index
\begin{equation}
\kappa_N \equiv \frac{\rho}{p}\frac{d p}{d \rho}\, ,
\end{equation}
that approximates $\kappa$ for low pressures, respectively.
Unfortunately, the value of $\kappa$ is not in the
preferred range $4/3-5/3$ that is considered to be physically
acceptable in the case of compact neutron stars or white dwarfs.

As we have seen in subsection\,\ref{constp}, the surfaces of constant
pressure, $\mathcal{S}_r$, are determined by the relation $r=\bar
r+\xi_0+\xi_2 P_2(\cos\theta)$, where $\xi_2$ is given by
(\ref{xi2}). These surfaces are oblate iff $k_2(\bar r)+\xi_2/\bar
r<0$. In terms of the functions $\nu$, $\beta$, $\zeta$, $\hat h$,
$\hat k$ and $\omega$ the expression $k_2(\bar r)+\xi_2/\bar r$ can be
written as
\begin{equation}
k_2(\bar r)+\xi_2/\bar r=e^{-2\nu}\left(r^4\hat k - r^2\hat h -\frac{(3\hat
  h +\omega^2)(1+\beta r^2)}{3(\zeta-\beta)}\right) \, . 
\end{equation}
Using the Taylor expansion around the centre, as it was made in
subsection\,\ref{sec:series}, one obtains in the limit $\bar r
\rightarrow 0$
\begin{equation}
\lim_{\bar r \rightarrow 0} k_2(\bar r)+\xi_2/\bar r=-\frac{1}{3} e^{-2\nu(0)}(\omega_0^2+3
h_1)/(z_1-b_1)\ .
\end{equation}
Note that $e^{\nu (0)}$ may be fixed to 1 since its value only
corresponds to a rescaling of the time-coordinate.  Since $z_1>0$ and
$b_1<0$ for realistic configurations we see that close to the centre
the surfaces of constant pressure are oblate iff $h_1>-\omega_0^2/3$.
In Table\,\ref{table:data} the values of $a\equiv\left(k_2(\bar
r)+\xi_2/\bar r\right)|_{\bar r=r_1}$ and $b\equiv\left(k_2(\bar
r)+\xi_2/\bar r\right)|_{\bar r=0}$ are also indicated for several
configurations. In virtue of the negative signs of these parameters
the constant pressure surfaces are all oblate for these
configurations.

In Table\,\ref{table:data} the central pressure $p_{0c}=2z_1$, the
radius of the zero pressure surface $r_1$, the shape at the zero
pressure surface, the shape close to centre, the value of $c_1$, the
maximal speed of sound $v_s^2$, the zeroth order density at the
zero pressure surface and the adiabatic index 
are given for a sequence of solutions with $q_1=0$.
\begin{table*} [h!]
{\small 
	  \centering
		\begin{tabular}{|l|l|l|l|l|l|l|l|l|l|} \hline   
      $h_1$ & $p_{0c}=2 z_1$ & $r_1$& $a$ & $b$ & $c_1$ & $v_s^2$  & $\rho_0(r_1)$ & $\kappa$ & $\kappa_N$ \\ \hline
			0.006 & 0.2184 & 0.264& -0.0086 &-0.00841 & $1.13\cdot10^3$ & 0.36  & 5.3021 & 10.3 & 9.9 \\ \hline
			0.007 & 0.4480 & 0.367& -0.0089 &-0.00844 & 46.675 & 0.44  & 4.7244 & 6.3 & 5.9 \\ \hline
			0.008  & 0.6896 & 0.441& -0.0091 &-0.00843 & 8.165 & 0.54  & 4.2416 & 5.2 & 4.7 \\ \hline
			0.009 & 0.9440 & 0.500& -0.0093 &-0.00839 & 2.556 & 0.64  & 3.8318 & 4.7 & 4.1 \\ \hline
			0.01 & 1.210 & 0.548& -0.0095 & -0.00831& 1.105 & 0.74  & 3.485 & 4.4 & 3.7 \\ \hline
			0.011 & 1.4872 & 0.587& -0.0098 &-0.00822 & 0.584 & 0.85  & 3.1898 & 4.3 & 3.4 \\ \hline
			0.012 & 1.7762 & 0.621& -0.0100 &-0.00812 & 0.352 & 0.97  & 2.9324 & 4.2 & 3.3 \\ \hline
			0.015 & 2.6978 & 0.700& -0.0109 &-0.007805 & 0.1259 & 1.35  & 2.3441 & 4.4 & 3.0 \\ \hline
			0.02 & 4.3666 & 0.786 & -0.0126 &-0.00733 & 0.0514 & 2.05  & 1.7335 & 4.9 & 2.8 \\ \hline
			0.05 & 15.2 & 1.029& -0.0254 &-0.00620 & 0.0193 & 6.76  & 0.66428 & 9.4 & 2.7 \\ \hline
			0.1 & 32.8 & 1.190 & -0.0449 & -0.00594& 0.0210 & 14.95  & 0.42164 & 17.7 & 2.7 \\ \hline
			1 & 312.8 & 1.636& -0.2297 &-0.00637 & 0.0503 & 164.58  & 0.21398 & 168 & 3.2 \\ \hline
		\end{tabular}
      \caption{\small The central pressure $p_{0c}=2z_1$, the radius of
	the zero pressure surface $r_1$, the shape of zero pressure surface
($a\equiv k_2+\xi_2/r_1$ is negative for an oblate
	configuration), the shape of the constant pressure surfaces close to centre given
by $b\equiv (k_2+\xi_2/\bar r)\vert_{\bar r=0}$, the value of the constant
	$c_1$, the maximal speed of sound $v_s^2$, the
zeroth order energy density at the matching surface and, finally, the central values of the adiabatic
	indices $\kappa=\frac{p+\rho}{p}\frac{dp}{d\rho}$ and $\kappa_N=\frac{\rho}{p}\frac{dp}{d\rho}$
            are shown for some configurations which can be matched to an asymptotically
	flat exterior (with $q_1=0$). Although for all listed configurations the central density
	$\rho_{0c}=6=-6b_1$ any central density can be obtained using the rescaling freedom (\ref{scale1}). }
\label{table:data}
}
\end{table*}

\section{Conclusions}

The most important finding of this paper is that a subclass of slowly
rotating perfect fluid balls of Petrov type D can be matched to
asymptotically flat vacuum spacetimes and also that in general slowly
rotating perfect fluid balls can be matched to non-asymptotically flat
vacuum exteriors determined by equations (\ref{eq0vacuum}),
(\ref{eq1vacuum}) and (\ref{eqk2}). Our numerical results support the
conclusion that neither of the Petrov type D inner fluid solutions can
be matched to second order to the Kerr metric, which is in accordance
with the general expectation that the ellipsoidal shape of the
rotating fluid ball produces an extra contribution to the quadrupole
moment which should also be present in the corresponding quadrupole
moment of the external field \cite{Bertietal,Hernandez}.  It was also
found that there is a range in parameter space for which the value of
the central pressure is relatively low and the speed of sound is also
subluminal. The equation of state was also determined for various
solutions belonging to the investigated class. It is clear that the
equation of state cannot be polytropic since, in general, the energy
density does not vanish at the zero pressure surface. Nevertheless,
the equation of state can be approximated close to the centre by a
polytropic one. Unfortunately, the corresponding adiabatic index
$\kappa$ was found to take values out of the physically preferred range.

\end{document}